\documentclass[prd,twocolumn,amsmath,nobibnotes,nofootinbib]{revtex4}

\usepackage{bm} \usepackage{graphicx} \usepackage{epstopdf}
 \def\ba{\begin{eqnarray}} \def\ea{\end{eqnarray}}
\def\be{\begin{equation}} \def\ee{\end{equation}} \def\({\left(}
\def\){\right)} \def\[{\left[} \def\]{\right]} \def\<{\left<}
\def\>{\right>}

\def\ba{\begin{eqnarray}}
\def\ea{\end{eqnarray}}
\def\be{\begin{equation}}
\def\ee{\end{equation}}
\def\({\left(}
\def\){\right)}
\def\[{\left[}
\def\]{\right]}
\def\<{\left<}
\def\>{\right>}

\begin{document}

\title{A pedagogical explanation for the non-renormalizability of gravity.}
\date{\today}

\author{Assaf Shomer}
\email{shomer@scipp.ucsc.edu}
\affiliation{SCIPP, University of California, Santa Cruz, CA 95064, USA}

\begin{abstract}
We present a short and intuitive argument explaining why gravity
is non-renormalizable. The argument is based on black-hole
domination of the high energy spectrum of gravity and not on the
standard perturbative irrelevance of the gravitational coupling.
This is a pedagogical note, containing textbook material that is
widely appreciated by experts and is by no means original.
\end{abstract}

\maketitle

\section{Introduction}

In this note we present what we perceive to be an intuitive
explanation of the familiar fact that gravity is not a
renormalizable quantum field theory. Since this is somehow the
place where gravity and quantum mechanics clash, we felt that even
though this is by now textbook material, some people (particularly
students and postdocs\footnote{We were first exposed to this point
of view through conversations with Tom Banks. The motivation for
writing this note is the author's feeling that he missed this
simple, yet central theme in his own education.}), may benefit
from its exposition in the way done in this note.

The crux of the argument, which appears e.g. in \cite{ab}, is one
line long and here it is: {\bf The very-high energy spectrum of
any $d$-dimensional quantum field theory is that of a
$d$-dimensional conformal field theory. This is not true for
gravity.} The rest of this note\footnote{After receiving feedback
to the first version of this note we feel it is important to
emphasize that the argument advocated here {\it is not} the
perturbative argument about the irrelevance of the gravitational
coupling, which can be circumvented along the lines of Weinberg's
{\it Asymptotic Safety.} The argument we use builds on a sharp
contradiction between the density of states in gravity, deduced
via the Bekenstein-Hawking entropy formula for black holes, and
the density of states in {\it any} renormalizable quantum field
theory. Even though Wilsonian renormalization is crucial to build
the argument (and in fact takes up most of the text), the reader
is advised to read through to the last section where the black
hole argument is presented. } is devoted to explaining this short
claim. We felt that real understanding had to go through a crash
course in Wilsonian renormalization, explaining some facts about
conformal field theories and finally pointing out why as a simple
consequence of black hole thermodynamics gravity can not be a
renormalizable quantum field theory.

The structure of this note is the following. We start in section
\ref{secren} by giving a crash course in Wilsonian
renormalization. In section \ref{secgrav}\ we discuss the
perturbative irrelevance of the gravitational coupling which is
the basis for the standard non-renormalization argument. In
section \ref{seccft}\ we discuss some relevant aspects of the
density of states in conformal field theory. In section
\ref{secarg}\ we present the argument for the
non-renormalizability of gravity, based on black hole entropy
considerations. We conclude in section \ref{secconc}.

\section{Renormalization}\label{secren}

It turns out to be a fact of nature that the low energy behavior
of many systems is largely independent of the details of what goes
on in higher energy scales. For example, you need not worry about
Feynman diagrams when analyzing the physics of waves in water. You
need not even worry about atoms. The Navier-Stokes equations,
formulated in the approximation of a continuous medium, are
enough. The actual physical theory describing the molecules and
their interaction does creep in through determining various
coefficients in the low energy theory (e.g. the viscosity
coefficient). However, once this quantity has been measured one
can use the Navier-Stokes equation to make predictions about the
propagation of waves in water and not worry about the underlying
theory\footnote{Of course, this is only valid as long as one works
within the energy regime where it is the correct effective
description. If you wish to explode a dynamite stick in water, the
Navier-Stokes equation is no longer a good low energy effective
theory.}. Those coarse grained descriptions applicable at low
energies are called {\it low energy effective field-theories.}
Even though field theory applications in, say, condensed matter,
are sometimes referred to as being not-fundamental in contrast to,
say, the standard model, a more honest statement is that they are
both ``just" effective descriptions applicable at different energy
scales\footnote{Incidentally, one sometime hears that perhaps
gravity should not be quantized at all because it is not a
``fundamental'' theory but is derived from a deeper structure, E.g
\cite{Jacobson}. One reason why it seems this is not the case is
that the same argument will tell you that you should not attempt
to quantize the effective field theory describing the motions of
atoms in a lattice, giving rise to phonons. So if you believe in
the standard model quanta, then phonons and also gravitons are
equally there.}. The LHC, soon to be turned on at CERN, is
designed to probe the TEV scale where the standard model
effectiveness as a low energy description is expected to break
down.

The modern theory of renormalization and its application to
quantum field theory was pioneered by Wilson, Fisher, Kadanoff and
Wegner \cite{seminal}. This subject is covered in many textbooks,
e.g. \cite{books}. The treatment of the RG equations is modelled
after Polchinski's paper \cite{polchinski}.

\subsection{Regularization}

Naively, even the simplest quantum field theories (QFT) are
useless because the answer to almost any calculation is infinite.
A standard example is the 1-loop correction to the mass in scalar
$\lambda\phi^3$
\begin{equation}\label{ems}
\begin{split}
&\Delta m^2=\frac{\lambda^2}{2}\int \frac{d^4p}{(2\pi)^4}\frac{1}{(p^2+m^2)((p+q)^2+m^2)}=\\
=&finite+c\cdot\int^{\infty}\frac{dp}{p}=\infty
\end{split}
\end{equation} for some constant $c$.
The reason we got this meaningless answer was the insistence on
integrating all the way to infinity in momentum space. This does
not make sense physically because our quantum field theory
description is bound to break at some point, if not sooner then
definitely\footnote{Even in free quantum field theories there is
an harmonic oscillator $\phi(x)$ in each spacetime point. Strictly
within quantum field theory we can excite this degree of freedom
to arbitrarily high energies. But once we add gravity to the game
this can no longer be true. The reason is that when enough energy
density is concentrated in a small region in space it will
collapse to form a black hole.} at the Planck scale ($p\sim
M_{Planck})$. One way to make sense of the theory is to introduce
a {\it high energy cut off} scale $\Lambda$ where we think the
theory stops being valid and allow only for momenta smaller than
the cut-off to run in the loops. But having done that, we run into
trouble with quantum mechanics, because usually, the regularized
theory is no longer unitary (since we arbitrarily removed part of
the phase space to which there was associated a non-zero
amplitude.) We therefore want to imagine a process of removing the
cutoff but leaving behind ``something that makes sense.'' A more
formal way of describing the goal is the following. We are probing
the system at some energy scale $\Lambda_R$ (namely, incoming
momenta in Feynman graphs obey $p\leq\Lambda_R$) while keeping in
our calculations a UV\footnote{As usual, we sometime refer to high
energies (compared to some scale) as ``the UV" and to low energies
as ``the IR". It should be kept in mind that these concepts are
very context dependant. The same energy scale can be referred to
as the UV in one discussion and the IR in another.} cutoff
$\Lambda$ ( $\Lambda\gg \Lambda_R$ because at the end of the day
we want to send $\Lambda\rightarrow\infty.$) If we can make all
physical observables at $\Lambda_R$ {\it independent} of $\Lambda$
then we can safely take $\Lambda\rightarrow\infty.$ In practice,
it is convenient to parameterize the energy scale using an RG
``time'' parameter flowing towards lower and lower energies
\begin{equation}\label{rgt}
\Lambda(t)=\Lambda(0)e^{-t}
\end{equation} and demand that {\it changing the cutoff $\Lambda$
leaves the partition function\footnote{With an appropriate restriction on incoming momenta $p\leq\Lambda_R$.} invariant.}
\begin{equation}\label{dtzisz}
\partial_tZ[J]=0.
\end{equation}
Let us see what the consequences of such an analysis are.

\subsection{The Renormalization Group (RG) equations}\label{deriv}

\begin{figure}
\includegraphics[width=5 cm,height=3.33 cm,bb=95 3 326 145]{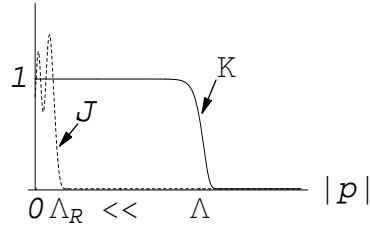}
\caption{The linear source term is constrained to $J(p)=0$ for
$p>\Lambda_R$ so as to only excite Green functions with low
energy. The quadratic term contains a smooth momentum cutoff
function $\mathcal{K}(\frac{p^2}{\Lambda^2})$ with the property
that $\mathcal{K}=1$ for $P<\Lambda$ and then falls off smoothly
(and exponentially fast) to zero for $P\geq\Lambda$.}
\label{fig-kj}
\end{figure}

We will model here the partition function of a quantum field
theory by
\begin{equation}
Z[J]=I[J]/I[0] {\rm\ \ where\ }
I[J]=\int[d\phi]e^{-\bigl(\mathcal{S}_0+\mathcal{S}_I+\mathcal{S}_J\bigr)}.
\end{equation} The action splits into the (linear) source term, the (quadratic) kinetic term and a general (polynomial) interaction term that we write in momentum space as
\begin{figure}
 \centering
 \includegraphics[width=6 cm,height=1.5 cm,bb=90 3 322 64]{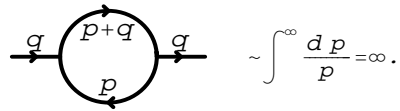}
\caption{The 1-loop correction to the mass in $\phi^3$ scalar field theory is naively divergent.}\label{fig-1loop}
\end{figure}
\begin{equation}\label{defes}
\begin{split}
\mathcal{S}_J&=\int\frac{d^4p}{(2\pi)^4}J(p,\Lambda_R)\phi(-p)\\ \mathcal{S}_0&=\frac{1}{2}\int\frac{d^4p}{(2\pi)^4}\phi(p)\phi(-p)\frac{\Delta(p)}{\mathcal{K}(\frac{p^2}{\Lambda^2})}\\
\mathcal{S}_I&=\sum_{n=3}^{\infty}\int\frac{d^4p_1}{(2\pi)^4}\cdots\frac{d^4p_n}{(2\pi)^4}\delta^4\bigl(\sum p_i\bigr)\\ &\qquad\qquad g_n(p_1,\cdots,p_n;\Lambda)\phi(p_1)\cdots\phi(p_n).\\
\end{split}
\end{equation} where $\Delta$ is the inverse propagator in momentum space\footnote{E.g. for a massive scalar field $\Delta=p^2+m^2$.} and we include a source function $J(p)$ and a smooth cutoff function $\mathcal{K}$ with specific properties that are explained in the caption of Fig. \ref{fig-kj}.

We now demand that the partition function remains invariant upon
an infinitesimal change of the high energy cutoff scale $\Lambda$
(Eq. \ref{dtzisz})
\begin{equation}\label{dennum}
\partial_t Z[J]=\frac{1}{I^2[0]}\Bigl(\partial_tI[J] \cdot I[0]-I[J]\cdot \partial_tI[0]\Bigr)=0.
\end{equation}
We begin by computing the scale derivative of $I[J]$
\begin{equation}\label{basice}
\partial_t I[J]=\int[d\phi]\Bigl\{\partial_t\bigl(e^{-\mathcal{S}_0-\mathcal{S}_J}\bigr)e^{-\mathcal{S}_I}+e^{-\mathcal{S}_0-\mathcal{S}_J}\partial_t\bigl(e^{-\mathcal{S}_I}\bigr) \Bigr\}
\end{equation}
We now use the definition of functional derivatives $\delta\phi(q)/\delta\phi(p)=\delta^4(p-q)$ to write
\begin{equation}
\begin{split}
&\int\frac{d^4p}{(2\pi)^4}\frac{\delta}{\delta\phi(p)}\frac{\delta}{\delta\phi(-p)}e^{-(\mathcal{S}_0+\mathcal{S}_J)}=\\ &\int\frac{d^4p}{(2\pi)^4}\frac{\delta}{\delta\phi(p)}\Bigl(-[\phi\frac{\Delta}{\mathcal{K}}+J](p)e^{-(\mathcal{S}_0+\mathcal{S}_J)}\Bigr)=\\
&\Bigl[-\delta^4(0)\cdot\int\frac{d^4p}{(2\pi)^4}\frac{\Delta}{\mathcal{K}}(p)\ +\\ &\int\frac{d^4p}{(2\pi)^4}\bigl[\phi\frac{\Delta}{\mathcal{K}}+J\bigr](p)\cdot\bigl[\phi\frac{\Delta}{\mathcal{K}}+J\bigr](-p)\Bigr]e^{-(\mathcal{S}_0+\mathcal{S}_J)}
\end{split}
\end{equation}
The first term includes an infinite factor corresponding to the
volume of spacetime\footnote{Physically, this term is related to
the renormalization of the vacuum energy.}, which will be
cancelled by an identical but opposite contribution from the
denominator. It now follows from the properties of $J,\mathcal{K}$
explained in Fig. \ref{fig-kj}\ that
\begin{itemize}
\item (a) $\partial_t\mathcal{K}\cdot J=0$ because they have disjoint support,
\item (b) $\partial_t J(p)=0$ because $J$ depends only on $\Lambda_R..$
\end{itemize} Using property $(a)$ the terms that include $J$ vanish and it follows that
\begin{equation}
\begin{split}
&\int\frac{d^4p}{(2\pi)^4}\frac{\partial_t\mathcal{K}}{\Delta}\frac{\delta}{\delta\phi(p)}\frac{\delta}{\delta\phi(-p)}e^{-(\mathcal{S}_0+\mathcal{S}_J)}=
\\ \Bigl[&\int\frac{d^4p}{(2\pi)^4}\frac{\Delta\partial_t\mathcal{K}}{\mathcal{K}^2}\phi(p)\phi(-p)- V\cdot\int\frac{d^4p}{(2\pi)^4}\frac{\partial_t\mathcal{K}}{\mathcal{K}}\Bigr] e^{-(\mathcal{S}_0+\mathcal{S}_J)}.
\end{split}
\end{equation}
Therefore, using property $(b)$ and the fact that $\mathcal{S}_0$
depends on the cutoff only through the function $\mathcal{K}$
\begin{equation}\label{thtv}
\begin{split}
&\partial_t \bigl(e^{-(\mathcal{S}_0+\mathcal{S}_J)}\bigr)=\\-&\frac{1}{2}\int\frac{d^4p}{(2\pi)^4}\phi(p)\phi(-p)\Delta\cdot\partial_t\bigl( \frac{1}{\mathcal{K}}\bigr)e^{-(\mathcal{S}_0+\mathcal{S}_J)}=\\&\frac{1}{2}\int\frac{d^4p}{(2\pi)^4}\frac{\partial_t\mathcal{K}}{\Delta}\frac{\delta}{\delta\phi(p)}\frac{\delta}{\delta\phi(-p)}e^{-(\mathcal{S}_0+\mathcal{S}_J)}+V\cdot A
\end{split}
\end{equation} where for brevity we call the c-number term with the explicit volume dependence $V\cdot A$.
So we have derived that
\begin{equation}\label{waiz}
\begin{split}
&\partial_tI[J]=\int[d\phi]\Bigl\{ \frac{1}{2}\int\frac{d^4p}{(2\pi)^4}\frac{\partial_t\mathcal{K}}{\Delta}\frac{\delta}{\delta\phi(p)}\frac{\delta}{\delta\phi(-p)}e^{-(\mathcal{S}_0+\mathcal{S}_J)}\\
&+V\cdot A\bigr)e^{-\mathcal{S}_I}+e^{-\mathcal{S}_0-\mathcal{S}_J}\partial_t\bigl(e^{-\mathcal{S}_I}\bigr) \Bigr\}
\end{split}
\end{equation}

The c-number term $V\cdot A$ is cancelled between the two terms in
Eq. \ref{dennum}\ because it can be pulled out of the path
integral. The remaining two terms do not cancel each other but
inspecting Eq. \ref{waiz}\ we see that they can both be set to
zero if we demand
\begin{equation}\label{waix}
\begin{split}
&\int[d\phi]\Bigl\{ \frac{1}{2}\int\frac{d^4p}{(2\pi)^4}\frac{\partial_t\mathcal{K}}{\Delta}\frac{\delta}{\delta\phi(p)}\frac{\delta}{\delta\phi(-p)}e^{-(\mathcal{S}_0+\mathcal{S}_J)}\bigr)e^{-\mathcal{S}_I}\\+&e^{-\mathcal{S}_0-\mathcal{S}_J}\partial_t\bigl(e^{-\mathcal{S}_I}\bigr) \Bigr\}=0
\end{split}
\end{equation}
Integrating twice by parts gives
\begin{equation}\label{dtsi}
\boxed{\partial_t\bigl(e^{-S_I}\bigr)=-\frac{1}{2}\int\frac{d^4p}{(2\pi)^4}\frac{\partial_t\mathcal{K}}{\Delta}\frac{\delta}{\delta\phi(p)}\frac{\delta}{\delta\phi(-p)}e^{-\mathcal{S}_I}}.
\end{equation}
Eq. \ref{dtsi}\ describes the infinitesimal change of the interaction Lagrangian upon changing the UV cutoff $\Lambda.$ This dependence of the coupling constants on the cutoff is called  {\it the $RG$ flow.} The procedure of decreasing the cutoff on $|p|$ infinitesimally from $\Lambda$ to $\Lambda-\delta\Lambda$ is called {\it integrating out a ``momentum shell''}. Notice that {\it  no infinities are encountered} because all the momentum integrals are done in a finite (actually infinitesimal) range.

\subsection{Asymptotics of the RG flow - fixed points }

Eq. \ref{dtsi}\ is a generalized version of the Heat equation
\begin{equation} \label{heat}
\partial_t\mathcal{X}=-\nabla^2\mathcal{X}
\end{equation} to a space of functions with a non-trivial positive definite bilinear form $\frac{\partial_t\mathcal{K}}{\Delta}$, so that the Laplacian is given by $\nabla^2=\frac{1}{2}\int\frac{d^4p}{(2\pi)^4}\frac{\partial_t\mathcal{K}}{\Delta}\frac{\delta}{\delta\phi(p)}\frac{\delta}{\delta\phi(-p)}$ and $\mathcal{X}=e^{-S_I}$. This kind of differential equation is called a ``Gradient Flow,'' and is characterized
by a potential function that {\it decreases} along the
flow\footnote{The potential function here is
$V(t)=\frac{1}{2}\int[d\phi]\mathcal{X}\nabla^2\mathcal{X}$, since
Eq. \ref{heat}\ becomes $\dot{\mathcal{X}}=-\delta V
/\delta\mathcal{X}$ and so $\partial_tV=\frac{\delta
V}{\delta\mathcal{X}}\dot{\mathcal{X}}=-(\frac{\delta
V}{\delta\mathcal{X}})^2$ which is negative definite. Actually,
there is a small error in this derivation because V depends on the
scale also through the definition of the generalized Laplacian.}.
Erroneously enough, this flow is {\it irreversible} and so the
name renormalization {\it Group} is unfortunate. For the
interaction Lagrangian itself Eq. \ref{dtsi}\ gives
\begin{equation}\label{efors}
\partial_t{\mathcal{S}_I}=(\nabla \mathcal{S}_I)\cdot(\nabla \mathcal{S}_I)-\nabla\cdot\nabla\mathcal{S}_I.
\end{equation} Expanding the coupling constants $g_n(p_1,\cdots,p_n;\Lambda)$ around zero momentum and labelling the whole set by some generalized index $I$ Eq. \ref{efors}\ can be summarized by the {\it $\beta$ function equations}
\begin{equation}\label{beta}
\boxed{\partial_t{g^I}=\beta^I_{\ J}g^J+\beta^I_{\ JK}g^Jg^K=\beta(g)}
\end{equation} which due to the form of Eq. \ref{efors}\ contains only linear and quadratic terms in the coupling constants $g_I.$

Let us pause for a quick recap. Starting from a theory defined at
the cutoff $\Lambda$ with a set of {\it bare couplings} we can get
the {\it same low energy physics} at some scale
$\Lambda_R<<\Lambda$ when we {\it integrate out} the momentum
shell $(\Lambda-\delta\Lambda,\Lambda)$ (i.e. lower the cutoff
from $\Lambda$ to $\Lambda-\delta\Lambda$) provided that we {\it
change the coupling constants as a function of the energy scale}
according to Eq. \ref{beta}. Keeping the scale $\Lambda_R$ of low
energy physics fixed and sending the high energy cutoff to
infinity corresponds to longer and longer RG-flows. We are thus
interested in the asymptotic $t\rightarrow\infty$ behavior of Eq.
\ref{dtsi}. The heart of the argument relies on the simple fact
that the asymptotic behavior of ``gradient flows'' is governed by
fixed points.
Those are the set of values for the parameters where the ``time''
derivative vanishes, so if we start from such a point, we stay there forever.

Since the positive function $V$ decreases with time, as $t\rightarrow\infty$ the limit
is either $V=0$, called {\it the trivial fixed point}, or some
$V=V_0>0$, called {\it a non-trivial fixed point.}
Non-trivial fixed points are the set of finite (or zero) values
for the couplings $g_{\star}$ that satisfy $\beta(g_{\star})=0$. The easiest
example of a non-trivial fixed point is the {\it Gaussian fixed
point}, or free field theory. By construction, non-trivial fixed
points correspond to {\it scale invariant field theories}. As a
matter of fact, most of the interesting Lorentz and scale
invariant theories, end up having an even larger symmetry group,
the {\it conformal group}. Such quantum field theories are called
Conformal Field Theories, or CFTs for short\footnote{We discuss
the conformal group a little more in the last section.}. The
trivial fixed point corresponds to an empty theory because when
couplings become infinitely large, a low energy experiment cannot excite any degree of freedom.

\subsection{Perturbative analysis near a fixed point}
It is very instructive to analyze the RG flow in the vicinity of a
fixed point.
\subsubsection{Classical analysis}
Linearizing the beta function equations (Eq. \ref{beta}) around a
fixed point $g_{\star}$, the {\it deviations} away from that fixed
point $D^I=g^I-g^I_{\star}$ satisfy $\partial_tD^I=\lambda^I_{\
J}D^J$ for some matrix $\lambda^I_{\ J}$. When $\lambda^I_{\ J}$
can be diagonalized, the eigenfunctions $\Delta^I$
satisfy\footnote{This analysis is applicable, of course, only as
long as $\Delta\ll 1.$}
\begin{equation}\label{elambdat}
\dot{\Delta}^I=\lambda^I\Delta^I\longrightarrow\boxed{\Delta^I(t)=\Delta^I(0)\cdot e^{\lambda^I t}}
\end{equation} There are thus 3 kinds of couplings:
\begin{equation}
\begin{split}
{\rm Relevant}\quad \lambda^I>0&\quad\Rightarrow\quad\lim_{t\rightarrow\infty}\Delta^I(t)=\infty\\
{\rm Marginal}\quad\lambda^I=0&\quad\Rightarrow\quad\Delta^I(t)=\Delta^I(0)\\
{\rm Irrelevant}\quad\lambda^I<0&\quad\Rightarrow\quad\lim_{t\rightarrow\infty}\Delta^I(t)=0.
\end{split}
\end{equation}

\subsubsection{Quantum Corrections}

There are two important features missed by the classical analysis.

\begin{itemize}
\item {\it Marginal operators:} Quantum corrections may show that
a coupling that seemed marginal classically is in fact relevant
(like the gauge coupling in QCD) or is in fact irrelevant (like in
scalar $\lambda\phi^4$ theory). Such couplings are called,
accordingly, {\it marginally relevant}, or {\it marginally
irrelevant}. A coupling that remains marginal quantum mechanically
(like the gauge coupling in $\mathcal{N}=4$ Supersymmetric
Yang-Mills) is called {\it truly marginal}. The existence of a
truly marginal coupling signals that there is a continuous family
of conformal field theories labelled by the value of a
dimensionless parameter\footnote{Even though from many
perspectives there are important differences between an
(ir)relevant and a marginally (ir)relevant coupling (e.g. the
latter only runs logarithmically with the energy scale) as far as
delivering the point aimed at in this note, we can ignore this
difference. So by ``(ir)relevant'' we will refer to (ir)relevant
or marginally (ir)relevant.}.

\item {\it Irrelevant operators:}
In the classical analysis they just die away. In fact we can solve the exact non-linear equation \ref{beta}\ which remains quadratic in the $\Delta$ basis
 \begin{equation}\label{delflow}
\partial_t\Delta^I=\lambda^I\Delta^I+\tilde{C}^I_{\ JK}\Delta^J\Delta^K.
\end{equation} The solution is
\begin{equation}\label{nonsol}
\Delta^I(t)=e^{\lambda^It}\Bigl(\Delta^I(0)+\int_0^tdse^{-\lambda^Is}\tilde{C}^I_{\ JK}\Delta^J(s)\Delta^K(s) \Bigr).
\end{equation}

Given a non-zero $\tilde{C}^I_{\ jk}$ with $I\in{\rm irrelevant}$
and $j,k\in{\rm relevant\ or\ marginal}$ {\it the asymptotic value
of the irrelevant parameter is completely determined by the value
of the relevant and marginal couplings.} It is only the initial
value $\Delta^I(0)$ of the irrelevant parameter which gets
exponentially washed away.

\end{itemize}

\begin{figure}
 \centering
 \includegraphics[width=6 cm,height=6 cm,bb=91 3 286 212]{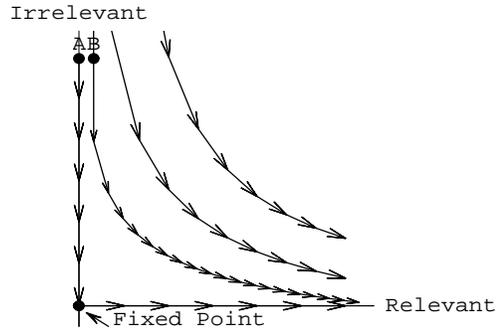}
\caption{The flow of coupling constants in the vicinity of a fixed
point with one relevant and one irrelevant direction. Point ``A"
represents a theory where all the relevant couplings are exactly
tuned to zero, which will therefore flow to the fixed point. Point
``B" represents a theory where at least one relevant coupling is
non-zero. Theory ``B" will approach the fixed point for a while
but then, as the relevant coupling starts to grow, it will be
``kicked''  out of the linear regime and will never hit the
CFT.}\label{fig-flow}
\end{figure}

Figure \ref{fig-flow}\ shows the qualitative features of such a
flow in a plane spanned by one relevant and one irrelevant
direction. Starting at the cutoff scale $\Lambda$ with a set of
couplings that flow so that near a fixed point there is a non-zero
value to any one of the relevant couplings, we will not hit that
fixed point. To make sure we hit the fixed point we need to tune
our initial couplings at the cutoff so that when we flow them to
the vicinity of that fixed point all the relevant couplings are
{\it exactly} zero. This set of constraints defines the {\it
critical manifold} for this fixed point. Tuning couplings to be on
the critical manifold guarantees approaching a scale invariant
behavior at low energies, which is by definition cutoff
independent. This may seem trivial but in fact, is very powerful.
The reason is that in practice, the vicinity of a fixed point will
contain a finite and small number of relevant couplings, but an
infinite number of irrelevant ones\footnote{E.g. in scalar field
theory in 4 dimensions, $:\phi^n:$ with $n>4$.}.  We can thus
guarantee having a well defined limit when
$\Lambda\rightarrow\infty$ by fiddling around only with a small
number of parameters.

We thus conclude that starting from any bare Lagrangian (UV) and
flowing to low energies (IR) using the RG equations, we
asymptotically end up either with an empty theory or a conformal
field theory. This gives us a first answer to the question raised
above, namely, a non trivial conformal field theory is ``something
that makes sense'' you can be left with as an effective low energy
description after removing a high energy cutoff.

\subsection{QFTs with a scale - dimensional transmutation}

If this was the end of the story it would not be very satisfying
because the theories we find in nature are not scale invariant. In
order to remove the cutoff and be left with a {\it
non}-scale-invariant quantum field theory we need to take a {\it
double scaling limit}. What this means, is that keeping a fixed
low energy scale $\Lambda_R$ we send the high energy cutoff
$\Lambda\rightarrow\infty,$ simultaneously sending the bare
relevant coupling constant closer and closer to the critical
surface, in such a way that the limit gives a theory with a finite
relevant coupling. To formulate this more precisely it is useful
to change the definition of the RG time (Eq. \ref{rgt}) so that
when $t=0$ we are at the low energy scale $\Lambda_R$ and sending
$\Lambda\rightarrow\infty$ corresponds to starting the flow at
increasingly negative times
\begin{equation}\label{rgt2}
\Lambda(t)=\Lambda_Re^{-t}.
\end{equation}
Using that, Eq. \ref{elambdat}\ gives
\begin{equation}\label{dimtrr}
\Delta(\Lambda_R)=\Bigl(\frac{\Lambda}{\Lambda_R}\Bigr)^\lambda\cdot\Delta(\Lambda)
\end{equation}
Therefore, we can keep a finite size\footnote{For Eq.
\ref{elambdat}\ to be valid, the deviation $\Delta$ must be
small.} for the {\it relevant} (i.e. $\lambda>0$) coupling
$\Delta(\Lambda_R)$ at low energies by sending
$\Lambda\rightarrow\infty$ and simultaneously sending the
deviation of the bare coupling away from the fixed point
$\Delta(\Lambda)\rightarrow 0$. More generally, starting with a
cutoff $\Lambda$ and $N$ relevant (and marginally relevant)
directions near a fixed point, we can take a double scaling limit
after which we are left with a finite length scale $\Lambda_R$ and
$N-1$ dimensionless parameters. One relevant coupling was
``traded\footnote{This means that we can solve for that relevant
parameter in terms of the energy scale $\Lambda_R$, and vice
versa.} for a scale'' giving this procedure the name {\it
dimensional transmutation}. The value of the other $N-1$ relevant
couplings encode exactly how we took the limit by fine tuning
closer and closer to the critical surface. The $N$ relevant
couplings are called {\it renormalized couplings.}

Dimensional transmutation is most familiar in marginally relevant
couplings. E.g. for a single marginally relevant parameter only
the quadratic piece from Eq. \ref{beta}\ can contribute so the
deviations from the fixed point obey $\partial_t\Delta=b\Delta^2$
(instead of Eq. \ref{elambdat}), and the solution is \footnote{We
changed the notations here and denote the deviations $\Delta$ by
$g$ so as to conform with familiar formulae where the fixed point
is free field theory. Note that solving for the scales in terms of
the couplings Eq. \ref{dimtrm}\ gives the  familiar
non-perturbative expression $\Lambda_R=\Lambda\cdot
e^{-(\frac{1}{b g(\Lambda)}-\frac{1}{b g(\Lambda_R)})}$.}
\begin{equation}\label{dimtrm}
\frac{1}{g(\Lambda_R)}=\frac{1}{g(\Lambda)}+b\ln\Bigl(\frac{\Lambda_R}{\Lambda}\Bigr),
\end{equation} where we can again keep $g(\Lambda_R)$ finite by sending $g(\Lambda)\rightarrow 0$
in a correlated way to that in which we send $\Lambda\rightarrow\infty.$

\subsection{Renormalizability and Universality}

While conformal field theories are few and far between in the
space of all possible quantum field theories, they are in this
sense the ``backbone'' defining all other quantum field theories.
A ``Renormalizable'' quantum field theory, describing the physics
at an energy scale $\Lambda_R$, is a perturbation of a conformal
field theory by some of its relevant operators with finite size
couplings at that scale. For example, massive scalar field theory
in 4 dimensions
$\mathcal{L}=\frac{1}{2}\bigl((\partial\phi)^2+m^2\phi^2\bigr)$ is
a perturbation by the relevant operator $\phi^2$ of the Gaussian
fixed point. Similarly, pure QCD is a perturbation of the Gaussian
fixed point $\mathcal{L}=\frac{1}{2}(\partial\mathcal{A})^2$ by
the marginally relevant coupling that follows from the gauge
invariant kinetic terms $\frac{1}{4}\mathcal{F}^2.$ The values of
the renormalizable couplings cannot be deduced from the bare
Lagrangian because their value depends on the arbitrary way one
chooses to remove the cutoff. Luckily, in many interesting cases
there is just a finite and small number of relevant parameters
which now encode {\it all} the physics of the theory at energy
scale $\Lambda_R$, including the values of the irrelevant
couplings, as was explained after Eq. \ref{nonsol}. Having
measured this finite number of couplings in the laboratory we can
make unique predictions about any experiment done in the energy
scale $\Lambda_R$ and remain ignorant about the information
encoded in (potentially infinitely many) irrelevant couplings.
This then gives a quantitative justification to the theme
advocated in the introduction, namely, the fantastic organizing
principle whereby phenomenon in one energy scale are ``shielded
off'' from the abyss of their own microscopic substructure.

In principle, it makes no difference what cutoff one
uses\footnote{In practice, calculational complexity may vary
dramatically using different cutoff schemes.} and how one chooses
to remove it. This important principle is called {\it
Universality}. For instance one can choose a sharp momentum cutoff
by refusing to continue the integration of momenta beyond
$\Lambda$, or a smooth cutoff as was done in this note. Other ways
include replacing space by a lattice with spacing $\Lambda^{-1}$,
or taking advantage of the analytic structure of the scattering
amplitudes by using dimensional regularization. These are
different ``Renormalization Schemes'' which are just different
ways of {\it parameterizing} the relevant and marginal couplings,
but the physics is only dictated by the fixed point around which
one is perturbing.

\subsection{Running back to high energies}

\begin{figure}
 \centering
 \includegraphics[width=3 cm,height=5 cm,bb=91 3 210 223]{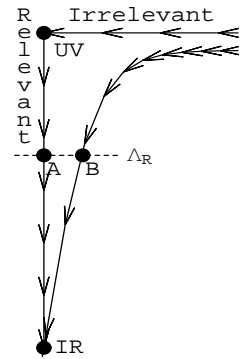}
\caption{Point (A) corresponds to a ``renormalizable'' QFT
describing the physics at an energy scale $\Lambda_R$. A UV fixed
point (e.g. $\mathcal{L}=\frac{1}{2}\bigl((\partial\phi)^2$) is
perturbed by a relevant operator (e.g. $\frac{1}{2}m^2\phi^2$).
Flowing further to lower energies it will eventually hit another
fixed point, the IR CFT (in this e.g. it is the trivial fixed
point $m=\infty$ as can be seen in Eq. \ref{ems}.). Point (B), on
the other hand, corresponds to a ``non-renormalizable'' QFT.
Adding a non-zero coupling to an irrelevant operator shoots us
away from the UV fixed point we naively thought we were
perturbing.}\label{fig-uvir}
\end{figure}

Given a renormalizable quantum field theory (point (A) in Fig.
\ref{fig-uvir}), we can run the RG flow {\it backwards} (i.e.
towards higher energies) where the relevant coupling vanishes.
Flowing further to low energies we will eventually hit another
fixed point, the IR conformal field theory. But what happens when
we attempt to add a finite size coupling to an irrelevant operator
at the scale $\Lambda_R$ (point (B) in Fig. \ref{fig-uvir}) and
then attempt to run the RG flow backwards\footnote{E.g. when
computing loop corrections in gravity.}? As can be seen in Fig.
\ref{fig-uvir}\ we get kicked off away from the UV fixed point,
the irrelevant couplings want to grow without bound. This is a
signal that we are doing something wrong. Going back to the
limiting process of removing the cutoff that defines a consistent
unitary quantum field theory we realize that we {\it cannot}
consistently remove the cutoff {\it and} retain a finite size for
an irrelevant coupling at low energies. Technically we can inspect
Eq. \ref{dimtrr}\ with $\lambda<0$ and observe that keep a finite
$\Delta(\Lambda_R),$ when $\Lambda\rightarrow\infty$ those
couplings need to grow without bound. The result of the
calculation is meaningless, exactly as we now understand it should
be, because we are trying to retrieve information that was lost in
the irreversible gradient flow during the limiting process that is
the basis for a consistent definition of any quantum field
theory\footnote{An amusing analogy is the formal infinite answers
to the illegal operation of dividing by zero. Multiplying any
number $x$ by zero is irreversible because the answer tells you
nothing about $x$. Dividing by zero is like insisting on
``undoing'' it, hence the ambiguous answer. Also here the correct
answer involves a more careful treatment of the limit, which
sometimes does and other times does not exist. }.

\subsection{A few remarks}

Before turning our attention to a particular Lagrangian, the
Einstein-Hilbert Lagrangian of General Relativity, we want to make
a few remarks:
\begin{itemize}

\item {\bf Path integrals and the Gaussian fixed point.} The
derivation of the path integral formula in quantum mechanics of a
massive particle involves chopping up the quantum evolution into
very short time intervals and inserting complete sets of states
between them. The transition amplitude between neighboring
spacetime points is
\begin{equation}
\begin{split}
&\langle x_n,t_n|x_{n-1},t_{n-1}\rangle\propto\\
&e^{\frac{i}{\hbar}\Delta t\Bigl(\bigl(\frac{m}{2}\bigr)\Bigl[\frac{(x_n-x_{n-1})}{\Delta t}\Bigr]^2-V\bigl(\frac{(x_n+x_{n-1})}{2}\bigr)\Bigr)}
\end{split}
\end{equation} which is dominated by free propagation in the limit $\Delta t\rightarrow0.$
Because in quantum mechanics short times are associated with high
energy (via the dimensionful constant $\hbar$), we can rephrase
the last sentence in the language of renormalization, where it
reads that the UV fixed point is free. More generally, only
quantum field theories that are perturbations of the Gaussian
fixed point can be accurately formulated as a path integral. It is
an interesting fact that all the quantum field theories useful to
date in describing physical reality share this property\footnote{Some
recent discussion of possible physical signatures of a non-trivial
fixed point are discussed e.g. in \cite{unptcl}.}.

\item{\bf Generality of RG equations.} Even though the RG
equations were derived here using a simple scalar field theory
path integral, this derivation is quite general. Its relevance
extends to non-trivial fixed points because to most of those one
can get by starting from a Gaussian fixed point and flowing down
the RG trajectory.

There are possible caveats to deriving the fixed point behavior as
a rigorous theorem. One has to do with theories that include
fermions, because arguments about the positivity of the inner
product would fail. Another has to do with gauge theories where it
may not be possible to implement a gauge invariant cutoff and
define the cutoff theory using a path integral. There are various
arguments suggesting that those are technical issues and the
general picture advocated here still holds (see e.g.
\cite{Morris1}).

\item {\bf Dimensionality of spacetime.} Renormalization theory
depends in an interesting way on the dimension of spacetime. E.g,
in the theory of a single scalar field in $d>6$ the only relevant
interaction of the Gaussian fixed point is a mass term. In 2
dimensions there are infinitely many marginal operators because
$\phi$ is of dimension zero. In 1 spacetime dimension (i.e.
quantum mechanics) there are infinitely many relevant operators.
In fact any function $V(x)$ is a relevant perturbation. This is
the reason that the problems of renormalization did not appear in
the old days of quantum mechanics.

\item {\bf Irrelevant interactions.} Here we chose to focus on a
rigorous removal of the cutoff and thus the inconsistency of
finite size irrelevant operators at low energies. However, in
practice, irrelevant operators are very useful as a tool in a low
energy effective field theory, as long as one is not pretending
that they are valid all the way to infinite energies. In fact, the
{\it dimensionful} couplings suppressing irrelevant operators are
an indication of the scale where the low energy effective
description breaks down. Marginally-irrelevant operators in
particular, since they run only logarithmically with the cutoff
(Eq. \ref{dimtrm}), may indicate new physics only at exponentially
high energies, where the validity of the theory is in any case expected to break down\footnote{A
good example is QED. The $U(1)$ gauge interaction is
marginally-irrelevant, but that does not really matter because the
energy scale where the theory is expected to break down is much
higher than, say, the electroweak symmetry breaking scale, where
QED is anyways incorporated into a larger theory.}.

\item {\bf Condensed matter.} The language, as well as much of the
intuition in the theory of renormalization in quantum field theory
is borrowed from the theory of second order phase transitions in
condensed matter physics.

\end{itemize}

\section{Gravity}\label{secgrav}

\subsection{The gravitational coupling is irrelevant}

The Einstein-Hilbert action
\begin{equation}\label{eh}
\mathcal{S}=\frac{1}{16\pi G_N}\int d^dx\sqrt{g}\mathcal{R}[g]
\end{equation} governing the dynamics of classical GR was arrived at via a symmetry principle, that of general coordinate invariance. In this respect it is very similar to the gauge theories one encounters in the standard model.
Using the defining property $ds^2=g_{\mu\nu}dx^{\mu}dx^{\nu}$ we see that the metric is dimensionless. Therefore $\Gamma\sim g^{-1}\partial g$ is of mass dimension 1 and the Lagrangian density $\sqrt{g}\mathcal{R}\sim \sqrt{g}\bigl(\Gamma^2+\partial\Gamma\bigr)$ is of mass dimension 2. Thus, the scaling dimension (with energy) of Newton's constant is
\begin{equation}\label{sdnc}
[G_N]_E=2-d
\end{equation}
It is customary to define $16\pi
G_N=2\kappa^2=(2\pi)^{d-3}l_d^{d-2}$ where $l_d$ is the
d-dimensional {\it Planck length}. This looks similar to a gauge
theory action $\frac{1}{4 g^2}\int d^dx\mathcal{F}^2$ where
$\kappa$ plays the role of the coupling constant. Indeed, we may
choose the dynamical variable to be the metric and expand around
flat space\footnote{It is immaterial which background metric we
choose because any smooth manifold is flat locally, so the short
distance behavior of GR is always like that of flat space.} by
defining $g_{\mu\nu}=\eta_{\mu\nu}+h_{\mu\nu}$, the first
non-vanishing contributions are schematically given by
\begin{equation}
\mathcal{S}\sim\frac{1}{2\kappa^2}\int d^dx \bigl[(\partial h)^2+(\partial h)^2h+\cdots\bigr]
\end{equation}
Rescaling $h=\kappa\tilde{h}$ in a way similar to the rescaling of
the gauge field in ordinary gauge theory, we get
\begin{equation}
\mathcal{S}\sim\frac{1}{2}\int d^dx \bigl[(\partial \tilde h)^2+\kappa(\partial \tilde h)^2\tilde h+\cdots\bigr]
\end{equation} which looks like a perturbation of the Gaussian fixed point $\int d^dx (\partial \tilde h)^2$.
Classically, this is fine, and means that gravity becomes free at
large distances. But quantum mechanically we observe that for
$d>2$ the {\it gravitational interaction is irrelevant}. The
previous discussion has shown that having a finite size for an
irrelevant parameter at some energy scale $\Lambda_R$ is
nonsensical quantum mechanically, unless we view the action (here
Eq. \ref{eh}) as a low energy field theory approximation of
something else. For instance, taking too literally a path integral
representation for the partition function in gravity, something
like $\int[dg]e^{-\int\sqrt{g}\mathcal{R}[g]},$ cannot make sense
because path integrals are by definition descriptions of quantum
field theories that are perturbations by relevant operators of
Gaussian fixed points.

Trying to solve the problem by avoiding Lagrangians all together
will not help. In a Hamiltonian formulation the same problem of
non-renormalizability, which technically appears in Lagrangian
theories as the need for infinitely many counter terms, creeps in
through the back door. There the Hamiltonian constraints, needed
for a consistent quantum mechanical treatment of a theory with a
gauge symmetry, do not close. Attempting to close the constraint
algebra  by adding the ``right hand side'' as an additional
constraint just leads to new terms and so on\footnote{In 2
spacetime dimensions the constraint algebra does close on the
Virasoro algebra, allowing for a consistent quantization of 2d
gravity. This is the starting point for string  perturbation
theory.}.

\subsection{Asymptotic Safety?}

At this point in the discussion it still appears to be possible
that there is another, non-trivial UV fixed point to which GR is a
perturbation by a relevant operator. This scenario was dubbed
``asymptotic safety'' by Weinberg \cite{as} and is reviewed e.g.
in \cite{revas} (for a recent attempt in this direction see e.g.
\cite{rizzo}). If that was the case then General Relativity is a
renormalizable quantum field theory and we were deceived by using
perturbation theory in the wrong variables. In the next sections
we explain why this can not be the case.

\section{Entropy in Conformal field theory}\label{seccft}

The conformal group can be defined as the group of coordinate
transformations preserving the form of the metric up to a scale
factor $g_{\mu\nu}(x)\rightarrow
g_{\mu\nu}^{\prime}(x^{\prime})=\Omega^2(x)g_{\mu\nu}(x).$ It is
an extension of the Poincare group that includes scale
transformation (dilatation) $x\rightarrow \lambda x.$ The
Conformal symmetry algebra in $d$ dimensional Lorentzian space is
isomorphic to $SO(2,d)$. Since the dilatation operator does not
commute with the momentum 4-vector $P^{\mu}$ we can no longer
characterize a representation of the conformal algebra by the mass
operator $M^2$ which is a Casimir of the Poincare group. For a
field that is an eigenfunction of the dilatation operator
$\phi(x)\rightarrow \lambda^{\Delta}\phi(\lambda x)$, the
eigenvalue\footnote{This is not to be confused with the eigenfunctions $\Delta^I$ that appeared in the discussion following Eq. \ref{elambdat}.} $\Delta$ is called the ``scaling dimension". Two
convenient ways to label states in a conformal field theory are to
decompose the $SO(2,d)$ into two different maximal subgroups:

\begin{itemize}
\item {1.} $SO(1,d-1)\times SO(1,1) \subset SO(2,d).$ In this case
the $SO(1,d-1)$ are Lorentz transformations in $d$ dimensions and
$SO(1,1)$ corresponds to dilatations. Fields are labelled by their
Lorentz representation and scaling dimension $\Delta$.

\item {2.} $SO(d)\times SO(2) \subset SO(2,d).$ This is a quantization on $R_{time}\times S^{d-1}$ (the temporal direction is the universal cover of the $SO(2)$ component). It is the natural choice in the operator formalism which makes a distinction between space and time. In this decomposition fields are labelled by their energy $E$ (the eigenvalue under $SO(2)$ translations, not to be confused with the time direction in $R^{1,d-1}$), and by spin quantum numbers associated with the sphere.

\end{itemize}

The two choices are related by the conformal symmetry. Starting from $R_{time}\times S^{d-1}$ and continuing to Euclidean signature the metric is $ds^2=dt_E^2+R^2d\Omega_{d-1}^2$ (with $R$ being the radius of the sphere). This is conformally equivalent to the metric on the Euclidean plane $ds^2=dr^2+r^2d\Omega_{d-1}^2$ if we set $t_E=\ln (r/R)$. From this relation we see that the Hamiltonian $\partial_t$ on $R_{time}\times S^{d-1}$ maps to the dilatation operator $r\partial_r$ on $R^{1,d-1}$. The relation between the quantum numbers is given by $\Delta=RE$.

Conformal invariance substantially simplifies some aspects of
finite temperature field theory. Since the zero temperature field
theory has no dimensionful scales in itself the temperature sets
the scale for all dimensionful quantities. Using dimensional
analysis and the extensivity of the energy and the entropy in $d$
spacetime dimensions it follows that in any conformal field theory
the energy (on $R_{time}\times S^{d-1}$) and entropy obey
\begin{equation}
S=a\cdot (RT)^{d-1},\qquad E=b\cdot R^{d-1}T^d
\end{equation} where $a,b$ are some numerical coefficients.
Solving for the entropy as a function of the energy one gets
\begin{equation}\label{cfts}
S\sim E^{\frac{d-1}{d}}.
\end{equation}
In the next section we compare this with the high energy spectrum
of gravity.

\section{Gravity is not a renormalizable quantum field
theory}\label{secarg}

\subsubsection{Zero cosmological constant}

As was thoroughly explained in the first part of this note, if
General Relativity was a renormalizable quantum field theory then
its extreme high energy behavior should be that of a conformal
field theory in the appropriate number of dimensions. However, our
experience with gravity has shown that once enough energy is
concentrated in a given region a black hole will form. As far as
our understanding goes, the high energy spectrum of GR is
dominated by black holes. More technically, it is expected that in
theories of gravity, black holes will provide the dominant
contribution to the large energy asymptotics of the density of
states as a function of the energy.

In asymptotically flat $d$ dimensional spacetime there is a black hole solution generalizing the Schwarzschild solution
\begin{equation}\label{grn}
\begin{split}
ds^2&=-f(r)dt^2+\frac{dr^2}{f(r)}+r^2 d\Omega_{d-2}^2, \\
f(r)&=1-\frac{\omega_{d-2}G_NM}{r^{d-3}},\\
\end{split}
\end{equation} where $\omega_n=\frac{16\pi}{n\cdot Vol(S^n)}$ and $M=(E)$ is the (ADM) energy of the black hole state.
The Horizon is thus at $r_H^{d-3}\sim G_NM$ and the entropy as a function of the energy is given by the Bekenstein-Hawking formula
\begin{equation}\label{bhs}
 \mathcal{S}=\frac{A}{4G_N}\sim \frac{r_H^{d-2}}{G_N}\sim \bigl(Ml_d\bigr)^{\frac{d-2}{d-3}}
\end{equation} where $A$ is the volume of the horizon and $l_d$ is the d-dimensional Planck length.
In this case $\mathcal{S}\sim E^{\frac{d-2}{d-3}}$ in
contrast\footnote{Note also that the exponent in Eq. \ref{bhs}\ is
bigger than one, rendering the specific heat negative which
reflects the thermodynamical instability of Schwarzschild black
holes. In contrast, the exponent in Eq. \ref{cfts}\ is smaller
than one, resulting in a positive specific heat.} to Eq.
\ref{cfts}. It therefore follows that the large energy asymptotics
of the density of states in a theory of gravity in asymptotically
flat spacetime is not that of any conformal field theory, and
therefore, it is not a renormalizable quantum field theory.

\subsubsection{Non-zero cosmological constant}

Asymptotically Anti de Sitter (AdS) space (the maximally symmetric
solution to Einstein's equations with a {\it negative}
cosmological constant $\Lambda=-3/R_{AdS}^2$) also admits black
hole solutions of the form $ds^2=-f(r)dt^2+\frac{dr^2}{f(r)}+r^2
d\Omega_{d-2}^2$ but in this case the function is
\begin{equation}
f(r)=1-\frac{\omega_{d-2}G_NM}{r^{d-3}}+\frac{r^2}{R_{AdS}^2}.
\end{equation} At distances much smaller than the AdS curvature radius $r\ll R_{AdS}$ this is almost the Schwarzschild solution (Eq .\ref{grn}), but asymptotically it is very different.
The horizon is at the larger solution to $f(r)=0$. Asymptotically $(r\gg R_{AdS})$ the horizon sits at $r_H\sim \bigl(\frac{G_N M}{|\Lambda|}\bigr)^{\frac{1}{d-1}}$ so that the entropy is
\begin{equation}\label{adsbh}
\mathcal{S}=\frac{A}{4G_N}\sim \Bigl(\frac{M R_{AdS}^2}{l_d}\Bigr)^{\frac{d-2}{d-1}}.
\end{equation}
Therefore, in this case $\mathcal{S}\sim E^{\frac{d-2}{d-1}}$.
Again, comparing this with Eq. \ref{cfts}\ the conclusion is that
gravity in $d$-dimensional AdS spacetime does not have the correct
density of states for a $d-$dimensions conformal field theory and
so cannot be a renormalizable quantum field theory in that sense.

However, there is a surprise here. Eq. \ref{adsbh}, gives the
correct dependence for a conformal field theory {\it but in one
less spacetime dimension}. This is a consequence of the AdS/CFT
correspondence \cite{adscft,wittenads,magoo}\ which claims a
complete equivalence, or duality, between quantum gravity in AdS
space and a conformal field theory (without gravity) defined, in a
precise sense that will not be explained here, on the boundary of
the AdS space. AdS/CFT itself is a manifestation of a modern
guiding principle in quantum gravity called ``holography''
\cite{thooft,susskind,bousso}.

The case of de Sitter space (the maximally symmetric solution to
Einstein's equations with a {\it positive} cosmological constant),
which seems to be the one relevant for our universe is much less
well understood. For once, it is not compatible with supersymmetry
which was a main tool in the advancement of our understanding in
other cases. There are reasons to suspect that quantum gravity in
dS space may not exist in its own right, perhaps only as a
meta-stable state in another
theory\cite{tom,finiteH,wittends,dscft,kklt}.

\section{Conclusions}\label{secconc}

In this note we tried to give a concise and hopefully intuitive
explanation to the fact that gravity is not a renormalizable
quantum field theory. The basic reason is that the asymptotic
density of states in gravity is dominated by black holes. This
leads to a behavior qualitatively different from all quantum field
theories.

A concern that was repeatedly raised after the appearance of the
first version of this note is that the black hole argument is an
artifact of the low energy approximation, so that ``strongly
coupled gravity" has no black-holes and therefore may still be
asymptotically safe. We believe this counter-argument does not
hold because the asymptotic safety scenario is based on the
assumption that gravity is a valid low energy approximation to
some putative local quantum field theory. Therefore at least in
its regime of validity it should be trusted. In particular it
should be trusted to describe the horizons of large black holes,
since as can be seen from Eq. \ref{grn} the {\it more} massive a
black hole is, the {\it lower} is the curvature at the horizon.
But the Bekenstein-Hawking formula tells us that the density of
states deduced from this valid approximation has a feature that is
clearly in contradiction with any quantum field theory. Therefore,
to argue against the non-renormalizability of gravity is really to
argue against the validity of the Bekenstein-Hawking formula,
which is an uphill battle. This is so much more so when taking
into account the AdS/CFT correspondence which verifies the
Bekenstein-Hawking entropy counting in the case of asymptotically
AdS space, and presents a clear counter example to the idea that
quantum gravity is a non-trivial fixed point.

It seems that gravity is a low energy effective field theory
description of something else that is not a quantum field theory.

\begin{acknowledgments}
We thank Riccardo Argurio, Stephen Adler, Tom Banks, Michael Dine,
Howard Haber and Asad Naqvi for comments on the manuscript.
Special thanks go to Tom Banks for introducing and patiently
explaining much of the content of this note.

\end{acknowledgments}


\end{document}